# A new class of topological insulators from I-III-IV half-Heusler compounds with strong band inversion strength

X. M. Zhang,[1] G. Z. Xu,[1] Y. Du,[1] E. K. Liu,[1] Z. Y. Liu,[2] W. H. Wang[1(a)] and G. H. Wu[1]

[1] Beijing National Laboratory for Condensed Matter Physics, Institute of Physics, Chinese Academy of Sciences, Beijing 100190, P. R. China

[2] State Key Laboratory of Metastable Material Sciences and Technology, Yanshan University, Qinhuangdao 066004, P. R. China

**Abstract:** In this paper, by first principle calculations, we investigate systematically the band topology of a new half-Heusler family with composition of I(A)-III(A)-IV(A). The results clearly show that many of the I-III-IV half-Heusler compounds are in fact promising to be topological insulator candidates. The characteristic feature of these new topological insulators is the naturally strong band inversion strength (up to -2eV) without containing heavy elements. Moreover, we found that both the band inversion strength and the topological insulating gap can be tailored through strain engineering, and therefore would be grown epitaxially in the form of films, and useful in spintronics and other applications.

# 1. Introduction

Topological insulators (TIs), as a new class of materials showing robust topological surface states inside the bulk insulating gap,[1-3] have attracted considerable research interests in condensed-matter physics.[4-8] Since the first two-dimensional (2D) TI with quantum spin-Hall effect was proposed in zinc-blende HgTe compound,[9,10] the TI family has grown with the discoveries of three-dimensional (3D) TIs in the tetradymite $Bi_2Se_3$ and $Bi_2Te_3$,[11,12] half-Heusler LaPtBi,[13,14] chalcopyrite $AuTlTe_2$,[15] layered honeycomb lattice $Na_2IrO_3$,[16] perovskite $YBiO_3$,[17] and also the wurtzite AgI and AuI.[18] So far, most of the existing TIs are heavy elements containing materials, since a strong spin-orbit coupling (SOC) effect related to heavy elements is assumed to be essential to the band inversion mechanism.[1,11] Very recently, however, Zhu *et.al* have reported that a topological phase transition can be achieved via strain engineering in light layered GaS and GaSe system.[19] The results suggest that the SOC may not necessary for inducing the topological band inversion. Actually, Feng *et.al* have predicted that the topological order of zinc-blende III-V and II-VI semiconductors can be turned by varying the lattice constant even with the SOC strength unchanged.[20] In the zinc-blende semiconductors, the 8 *sp* electrons (i.e. closed shell) configuration supplied by two main group elements is believed to be crucial to keep their phase stability and the semiconducting character as well.[21]

In this work, we theoretically demonstrate the possibility of exploring TIs in a new half-Heusler family with 8 valence electrons configuration, using the composition of I(A)-III(A)-IV(A) as a prototype system. The I-III-IV half-Heusler compounds are very different from the conventional half-Heusler TIs in: (i) The I-III-IV half-Heusler compounds are completely comprised by main group elements but contain no transition or tombarthite metals, and (ii) 8 valence electrons configuration (8 *sp*) is selected rather than the previous 18 electrons one (8 *sp* 10 *d*) to obtain an ideal semiconducting state. As a result, we found that there exist a large number of TI candidates in the I-III-IV half-Heusler compounds. Importantly, the new TIs can exhibit strong band inversion character but without containing heavy or rare earth elements. From application point of view, we also simulate experimental condition by performing strain engineering for some selected I-III-IV half-Heusler compounds.

## 2. Computational details

The band-structure calculations are performed using full-potential linearized augmented plane-wave method implemented in the WIEN2K package.[22,23] A converged ground state was obtained using 10 000 k points in the first Brillouin zone with $K_{max}* R_{MT} = 9.0$, where $R_{MT}$ represents the muffin-tin radius and $K_{max}$ is the maximum size of the reciprocal-lattice vectors. Wave functions and potentials inside the atomic sphere are expanded in spherical harmonics up to l=10 and 4, respectively. A combination of modified Becke-Johnson exchange potential and the correlation potential of the local-density approximation (MBJLDA) are used to obtain the band structures,[24] as it predicts band gaps and the band order with favorable accuracy similar to more expensive GW calculation.[25-27] Spin-orbit coupling is treated by means of the second variational procedure with the scalar-relativistic calculation as basis,[22] where states up to 10 Ry above the Fermi level are included in the basis expansion.

## 3. Results and discussions

Figure 1 (right plane) shows the crystal structure and the constituent elements for the I-III-IV half-Heusler compounds. Atoms from group-I(A), III(A) and IV(A) occupy the following three crystal sites of the cubic lattice: A(0, 0, 0), B(1/4, 1/4, 1/4), C(1/2, 1/2, 1/2). It has to be noted that, the actual lattice occupancy for the I-III-IV half-Heusler compounds cannot be taken for granted derive from the nomenclature and special care has to be taken to assign the atomic parameters correctly. With different atoms sitting the unique B site, there theoretically exist three inequivalent atomic arrangements for the half-Heusler structure as summarized in Fig. 1 (right plane). To determine the equilibrium lattice constants and testify the structure type, we respectively perform geometric optimizations for the I-III-IV half-Heusler compounds within the three structure types. The equilibrium energy as a function of the average nuclear charge <Z> is shown in Fig. 1 (left plane). The equilibrium energy of the I-III-IV compounds for structure type-I have been set as the zero point of the energy, and those for structure type-II and type-III are given as half-solid and empty shapes. Obviously, among the three structure types, the equilibrium energy of all the studied I-III-IV half-Heusler compounds for the type-III are in the highest energy level and indicate they are the least likely to follow such an atomic arrangement with the atoms from group-I(A) occupying the B site. While the calculated equilibrium energy for the other two types are very close. For the MSiAl (M=Na, K, Rb, Cs), NaGeGa and NaSnIn compounds, the equilibrium energy for structure type-II is lower, indicating form a $A^{I}B^{IV}C^{III}$ phase, similar to the well-known half-Heusler semiconductors LiAlSi and LiGaSi.[28,29] However, the other I-III-IV compounds are energy-preferred to form structure type-I with $A^{I}B^{III}C^{IV}$ atomic arrangement, which are consistent to the conventional half-Heusler system.

We next explore the band topology of the I-III-IV half-Heusler compounds, using KGaGe and RbSiAl as examples. In Fig. 2 (left plane), we show the topological band structures of (a) KGaGe and (b) RbSiAl compounds in their equilibrium states. The results show that both KGaGe in type-I and RbSiAl in type-II can form perfect HgTe/CdTe-like band structures. We thus only focus on the bands close to the Fermi level, which are the $s$-like $\Gamma_6$ bands and the $p$-like $\Gamma_7$ and $\Gamma_8$ ones. The band orders of KGaGe and RbSiAl are $\Gamma_8$ (red lines) →$\Gamma_7$ (blue lines) →$\Gamma_6$ (green lines) and $\Gamma_6$ →$\Gamma_8$ →$\Gamma_7$ from top to bottom, respectively. By considering of the band topology, KGaGe possesses a naturally inverted band order with the $\Gamma_6$ states sitting below the $\Gamma_8$ ones and guarantee its topologically nontrivial character; while RbSiAl exhibits a normal band order with $\Gamma_6$ states in a higher energy level, performing a trivial semiconducting behavior. To better understand the origins of corresponding bands in detail, we have drawn schematically in Fig. 2 (right plane) the possible hybridization between atoms for (a) KGaGe and (b) RbSiAl. We have taken into account the combining effects of chemical bonding, crystal field and SOC. And the resulting coupled hybridizing states are given as $s_\pm$, $p_{xy,\pm}$ and $p_{z,\pm}$, where "±" are the party labels. From the given hybridization diagram, we see that the $\Gamma_6$, $\Gamma_7$ and $\Gamma_8$ bands in the I-III-IV compounds most originate from the hybridizations of $s$ and $p_{xyz}$ orbitals between the atoms from III-IV zinc-blende sublattice. Our results are consistent with the description by Chadov et.al in half-Heusler XYZ system,[13] where the band structure near the Fermi level at the Γ point is determined mainly by the YZ zinc-blende sublattice.

As we discussed above, the band structures of the I-III-IV half-Heusler compounds are very similar to those of HgTe/CdTe and previous half-Heusler XYZ system as well.[9-14] Therefore, their band topology can be characterized by a function of $E_{BIS} = E_{\Gamma 6} - E_{\Gamma 8}$, namely, topological band inversion strength. In Fig. 3, we have summarized the $E_{BIS}$ as a function of (a) the lattice constant and (b) the average nuclear charge <Z> for the I-III-IV compounds. Here we show as three categories by different III-IV zinc-blende sublattice composition to discuss. It is clear that the MSiAl (M=Na, K, Rb, Cs) category of compounds all exhibit positive $E_{BIS}$, while the MGaGe (M=Na, K, Rb, Cs) and MInSn (M=Na, K, Rb, Cs) categories are more likely to show topological nontrivial behavior with negative $E_{BIS}$, except NaGeGa possess a zero value of $E_{BIS}$. Moreover, the most I-III-IV half-Heusler system shows strong band inversion strength (up to -2eV in CsInSn and CsGaGe), which can guarantee high robustness of the TI state upon external fluctuation.

Another phenomenon can be drawn from Fig. 3 (a) and (b) that, for the compounds with the same III-IV zinc-blende sublattice, the value of $E_{BIS}$ tends to decrease with the increase of the lattice constant [see Fig. 3 (a)] and the average nuclear charge <Z> [see Fig. 3 (b)]. According to the tight-bonding analysis by Feng *et.al* in Ref. 20 that, the topological band order in cubic semiconductors is mainly relevant to two factors: the coupling potentials of *s-s* and *p-p* bonds ($V_{ss}$ and $V_{pp}$), which are sensitive to the lattice constant; and the SOC strength ($\lambda_{SO}$) related to heavy elements. For the studied I-III-IV half-Heusler compounds, the SOC effect may be feeble since they contain no heavy elements (Z>60). To show the strength of band inversion purely induced by SOC, in the inset of Fig. 3 (a) we give out the energy discrepancy of $\Delta E_{BIS}= E_{BIS}^{SOC} - E_{BIS}^{W/S}$, where $E_{BIS}^{SOC}$ and $E_{BIS}^{W/S}$ represent the $E_{BIS}$ calculated with and without SOC. As anticipated, the SOC-induced $\Delta E_{BIS}$ for the MSiAl, MGaGe and MSnIn (M=Na, K, Rb, Cs) compounds are all very small with value lower than -0.01eV, -0.06eV and -0.16eV, respectively. Thus we argue that, among the above two factors, the former one, i.e., coupling potentials of *s-s* and *p-p* bonds, is more dominant to the topological band order in the I-III-IV half-Heusler compounds. From a further insight of the tight-bonding analysis in Ref. 20, the energy discrepancy of "$E_{BIS}^{SOC} - E_{BIS}^{W/S}$" and "$E_{\Gamma 7}- E_{\Gamma 8}$" are generally proportional to $1/2*\lambda_{SO}$ and $3/2*\lambda_{SO}$, respectively. In the inset of Fig. 3(b) we show the value of "$E_{\Gamma 7}- E_{\Gamma 8}$" for the I-III-IV compounds. Compared with the parameter "$E_{BIS}^{SOC} - E_{BIS}^{W/S}$" [see the insets of Fig.3 (a)], the "$E_{\Gamma 7}- E_{\Gamma 8}$" are triple in value, indicating the previous pioneering model surely well apply to the semiconductors in the I-III-IV half-Heusler compounds.

It has been widely accepted that, strain plays an important role in the band topology of TIs.[14,19,20,30,31] In the following, we will discuss the effects of two types of strain (hydrostatic and uniaxial) on reconstructing the band topology of the I-III-IV half-Heusler compounds. First, we show the effect of the hydrostatic strain, which can be achieved by varying the lattice constants with the cubic lattice retained. From the aspect of experimental condition, the hydrostatic strain may in some degree simulate external temperature or pressure fluctuations. In Fig. 4 (a) we show the $E_{BIS}$ of the NaGeGa compound changes on scaling the lattice constant with corresponding band structures given as insets. At the equilibrium lattice constant (6.23Å), NaGeGa exhibits a zero-$E_{BIS}$, i.e., $\Gamma_6$ and $\Gamma_8$ states are at the same energy level. While a small fluctuation to the lattice constant can significantly influence the band structure close to the Fermi level: the compression to the NaGeGa lattice leads to a trivial semiconducting phase with $\Gamma_6$ states situating above the Fermi level; whereas the expansion gives rise to a truly inverted band order between $\Gamma_6$ and $\Gamma_8$ states. With the increase of the lattice constant, the $E_{BIS}$ for the NaGeGa compound approximately linearly reduces. The reason can be briefly described as: with increasing the lattice constant, the coupling potentials $V_{ss}$ and $V_{pp}$ will reduce, which lowers the $\Gamma_6$ anti-bonding states and raises the $\Gamma_7$ bonding states, inducing the decrease of $E_{BIS}$ [20]. We can conclude that, similar as the conventional half-Heusler TIs, the topological band order of the I-III-IV half-Heusler compounds can be controlled by the hydrostatic strain, which may be helpful to the design of devices based on TI.

We next discuss the uniaxial strain, which provides at best the experimental condition for thin film growth. Here we simulate the epitaxial growth of RbGaGe compound on the semiconductor InSb substrate. It can be achieved by carrying out calculations using the lattice constant of InSb substrate (6.479Å, Ref.32) as that of the *ab*-plane of the uniaxial strained-RbGaGe, while leaving the *c*-axis unconstrained (free to relax). To determine the equilibrium value of *c*, a total energy minimization is performed and the results are shown in Fig. 4(b). It is found that the tetragonal-RbGaGe with c=6.948Å exhibits the lowest energy, and the inset of Fig. 4(b) gives out corresponding band structure. Unlike the zero-gap TIs in cubic system, the epitaxial RbGaGe can achieve a 'true' topological insulating state: with the inverted band order retained, the degeneracy of the $\Gamma_8$ states splits and opens up a globe gap of 88.3meV at the $\Gamma$ point. Our results suggest that the uniaxial strain encountered during epitaxial growth of films can induce topological transition from zero-gap semiconductors to topological insulator states.

## 4. Summary

In the present study, we show by fully relativistic first-principle calculations that, many of the I-III-IV half-Heusler compounds are promising to be topological insulator candidates with naturally inverted band order. The 8-electron configuration I-III-IV half-Heusler TIs feature with strong band inversion character but contain no heavy or rare-earth elements. It is also confirmed that, rather than the spin-orbit coupling factor, the coupling potentials of *s-s* and *p-p* bonds between the atoms from III-IV zinc-blend sublattice are dominant to the topological band order in our work. From application point of view, we simulate experimental conditions by performing the hydrostatic and the uniaxial strain for selected I-III-IV half-Heusler compounds. It is found that, the hydrostatic strain can continuously turn the band inversion strength; while the uniaxial strain is proved to be effective in driving the zero-gap semiconductor into a 'true' topological insulating state.

**Acknowledgements**

This work is supported by the National Basic Research Program of China (973 Program 2012CB619405) and National Natural Science Foundation of China (Grant Nos. 51071172, 51171207 and 51171206)

**Figure captions:**

FIG. 1. (right plane) The element composition, crystal structure and inequivalent site occupancy types for the I-III-IV half-Heusler compounds. (left plane) The equilibrium energy as a function of the average nuclear charge <Z> for the I-III-IV half-Heusler compounds. The equilibrium energy for structure type-I have been set as the zero point of the energy, and those for structure type-II and type-III are given as half-solid and empty shapes.

FIG. 2. The topological band structures and possible hybridization between atoms of (a) KGaGe and (b) RbSiAl compounds in their equilibrium states. KGaGe and RbSiAl are energy preferable to structure type-I and type-II, respectively.

FIG. 3. Band inversion strength ($E_{BIS}$) of the I-III-IV half-Heusler compounds as a function of (a) the lattice constant and (b) the average nuclear charge <Z>. In the middle it schematically shows that, a topological nontrivial phase features a negative $E_{BIS}$, while a trivial phase exhibits a positive $E_{BIS}$. The inset of (a) shows the energy discrepancy of calculated $E_{BIS}$ with and without SOC, and the inset of (b) shows the value of "$E_{\Gamma 7}$- $E_{\Gamma 8}$" calculated with SOC.

FIG. 4. (a) Band in inversion strength ($E_{BIS}$) as a function of the lattice constant for NaGeGa compound with corresponding band structures shown as insets. (b) Total energy as a function of the tetragonal lattice constant c, with the lattice constant of the *ab*-plane fixed as 6.479Å. The inset of (b) shows the band structure of RbGaGe with equilibrium *c* (6.948Å).

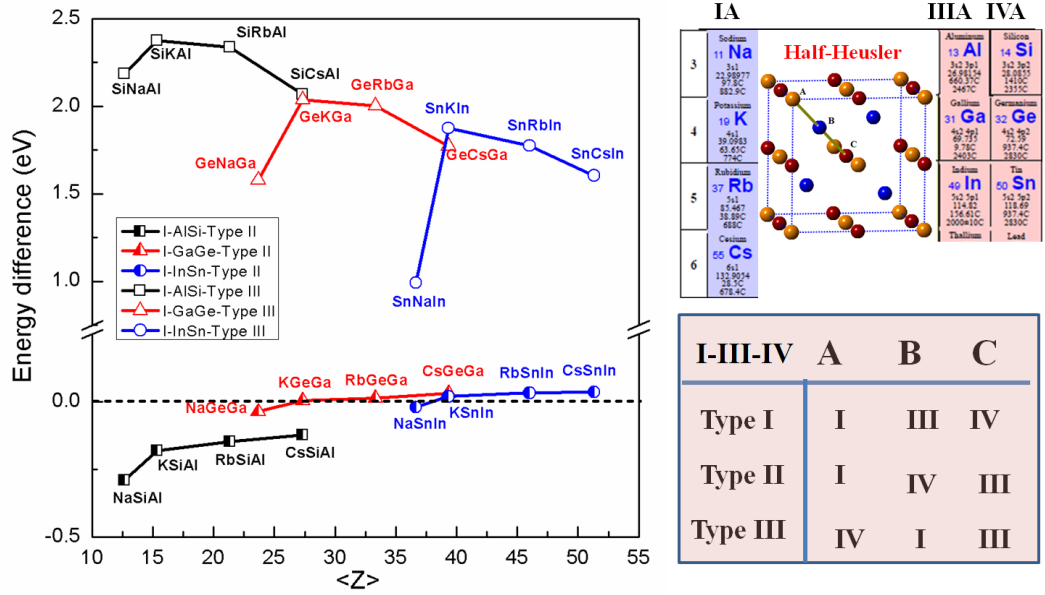

FIG. 1. Zhang et al.,

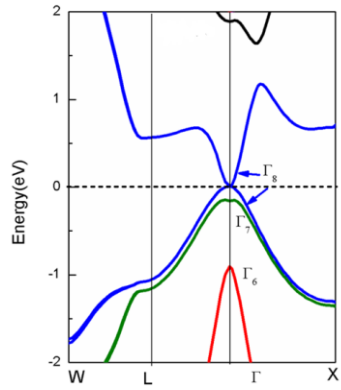
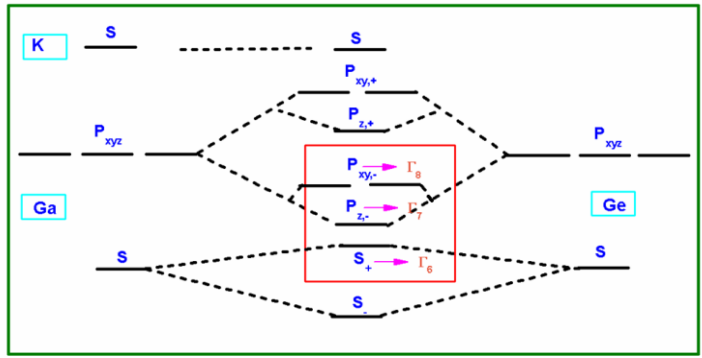
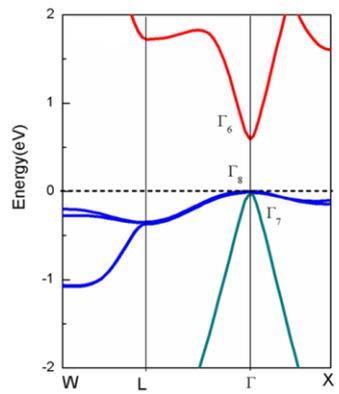
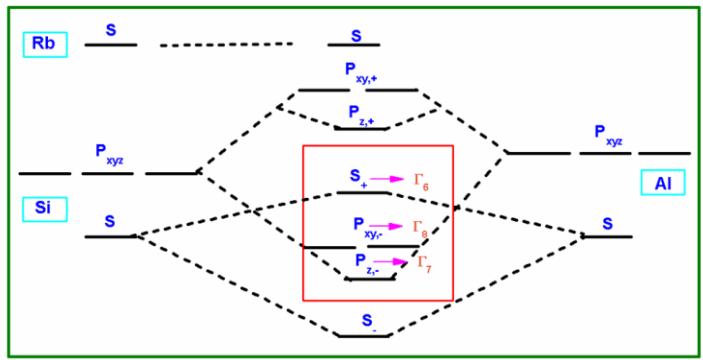

FIG. 2. Zhang et al.,

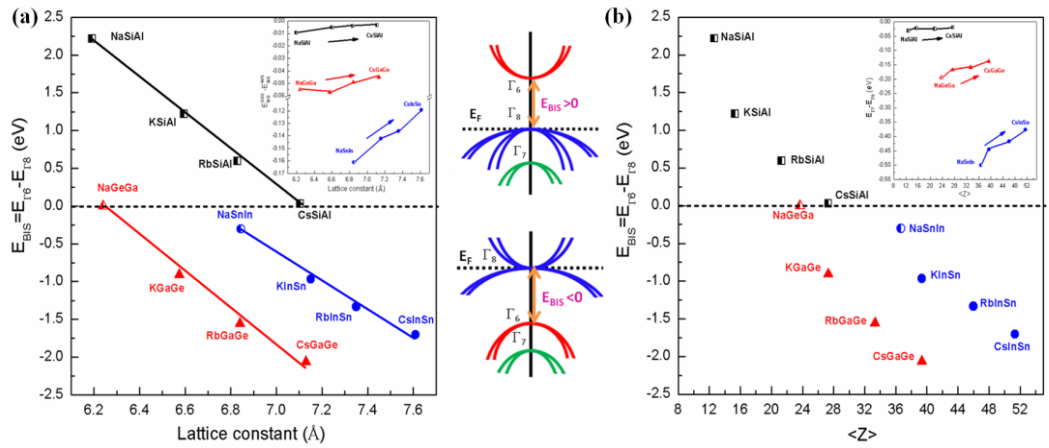

FIG. 3. Zhang et al.,

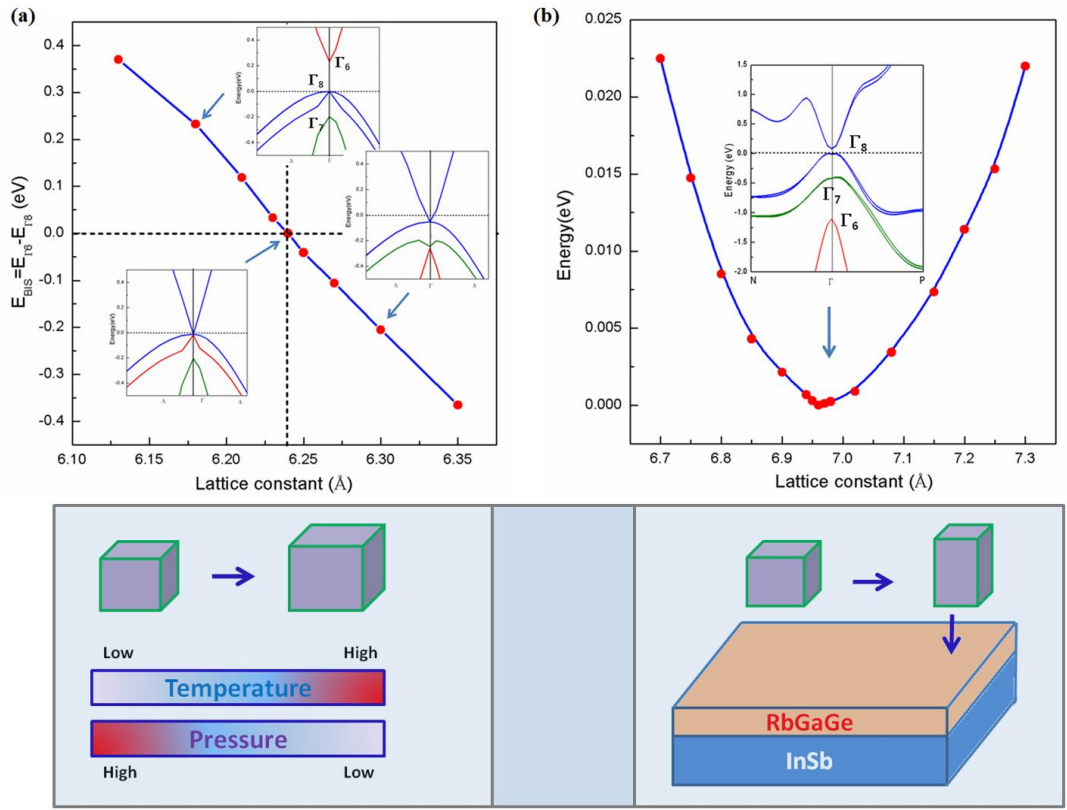

FIG. 4. Zhang et al.,